\documentstyle[epsfig,longtable]{aipproc}

\newcommand{\puncspace}{\ifmmode\,\else{\ifcat.\C{\if.\C\else%
\if,\C\else\if?\C\else\if:\C\else\if;\C\else\if-\C\else%
\if)\C\else\if/\C\else\if]\C\else\if'\C%
\else\space\fi\fi\fi\fi\fi\fi\fi\fi\fi\fi}%
\else\if\empty\C\else\if\space\C\else\space\fi\fi\fi}\fi}%
\newcommand{\SP}{\let\\=\empty\futurelet\C\puncspace}
\newcommand{\dash}{\hbox{--}}
\newcommand{\Hz}{{\hbox{Hz}}\SP}
\def\fu#1{\leavevmode\hbox{4U~#1}\SP}
\newcommand{\nonrot}{{\rm 0}}

\begin{document}

\title{Constraints on Neutron Star Masses and
Radii from Kilohertz QPOs}

\author{Frederick K. Lamb$^*$,
M. Coleman Miller$^{\dagger}$, and
Dimitrios Psaltis$^{\ddag}$}

\address{$^*$University of Illinois at
Urbana-Champaign, Department of Physics and
Department of Astronomy, 1110 W. Green St.,
Urbana, IL  61801\\
 $^{\dagger}$University of Chicago, Department
of Astronomy and Astrophysics, 5640 S. Ellis
Avenue, Chicago, IL  60637\\
 $^{\ddag}$Harvard-Smithsonian Center for
Astrophysics, 60 Garden St., Cambridge, MA
20218}

\maketitle

 \begin{abstract}
The frequencies of the highest-frequency
kilohertz QPOs recently discovered in some
sixteen neutron stars in low-mass X-ray binary
systems are most likely orbital frequencies.
 If so, these QPOs provide tight upper bounds on
the masses and radii of these neutron stars and
interesting new constraints on the equation of
state of neutron star matter. If the
frequency of a kilohertz QPO can be
established as the orbital frequency of the
innermost stable circular orbit, this would
confirm one of the key predictions of general
relativity in the strong-field regime. If the
spin frequency of the neutron star can also be
determined, the frequency of the QPO would fix
the mass of the neutron star for each assumed
equation of state.
 Here we describe how bounds on the stellar mass
and radius can be derived and how these bounds
are affected by the stellar spin. We also
discuss detection of the innermost stable
circular orbit.
 \end{abstract}

\section*{Introduction}

Determination of the equation of state of
neutron stars has been an important goal of
nuclear physics for more than two decades.
Progress toward this goal can be made by
establishing astrophysical constraints as well
as by improving our understanding of nuclear
forces. The discovery using the {\em Rossi X-ray
Timing Explorer\/} of highly coherent kilohertz
QPOs in the persistent emission of some sixteen
neutron stars in low-mass X-ray binaries (see
[1]) is likely to provide important new
constraints. In both the sonic-point
[2] and magnetospheric [3]
beat-frequency interpretations, the highest
frequencies observed are the orbital frequency
at the inner edge of the Keplerian disk flow (see
[4]). Such high orbital frequencies yield
interesting bounds on the masses and radii of
these neutron stars and interesting constraints
on the equation of state of neutron star matter.
 Here we describe how bounds on masses and radii
can be derived from the properties of the
kilohertz QPOs and how these bounds are affected
by stellar rotation. We also discuss detection
of the innermost stable circular orbit (ISCO).

\section*{Calculations}

Suppose that $\nu_{\rm QPO2}^\ast$ is the highest
Keplerian frequency observed from a given
neutron star and that the star is not rotating
(we use a superscript zero to indicate relations
that are valid only for a nonrotating star). The
radius of the star must be smaller than the
radius $R_{\rm orb}$ of the gas with orbital
frequency $\nu_{\rm QPO2}^\ast$, so the
representative point of the star in the
$R$,$M$ plane must lie to the left of the curve
$M^\nonrot(R_{\rm orb},\nu_{\rm QPO2}^\ast)$
that relates $R_{\rm orb}$, $\nu_{\rm
QPO2}^\ast$, and the mass of the star. In
addition, in order to produce a wave train with
tens of oscillations, the gas producing the QPO
must be outside the radius $R_{\rm ms}$ of the
marginally stable orbit (ISCO), so the
representative point must also lie below the
intersection of $M^\nonrot(R_{\rm orb},\nu_{\rm
QPO2}^\ast)$ with the curve $M^\nonrot(R_{\rm
ms})$ that relates $R_{\rm ms}$ and the mass of
the star.
 Figure~1a shows the allowed region of the
$R$,$M$ plane for $\nu_{\rm QPO2}^\ast =
1220~\Hz$. The maximum allowed mass and radius
are [2]
 \begin{equation}
  M^\nonrot_{\rm max} =
  2.2\,(1000~\Hz/\nu_{\rm QPO2}^\ast)\; M_\odot
  \quad{\rm and}\quad
  R^\nonrot_{\rm max} =
  19.5\,(1000~\Hz/\nu_{\rm QPO2}^\ast)\;{\rm km.}
 \end{equation}
 Figure~1b compares the $M$-$R$ relations given
by five equations of state with the allowed
regions of the radius-mass plane for three
values of $\nu_{\rm QPO2}^\ast$.

 \begin{figure}
\centerline{\epsfig{file=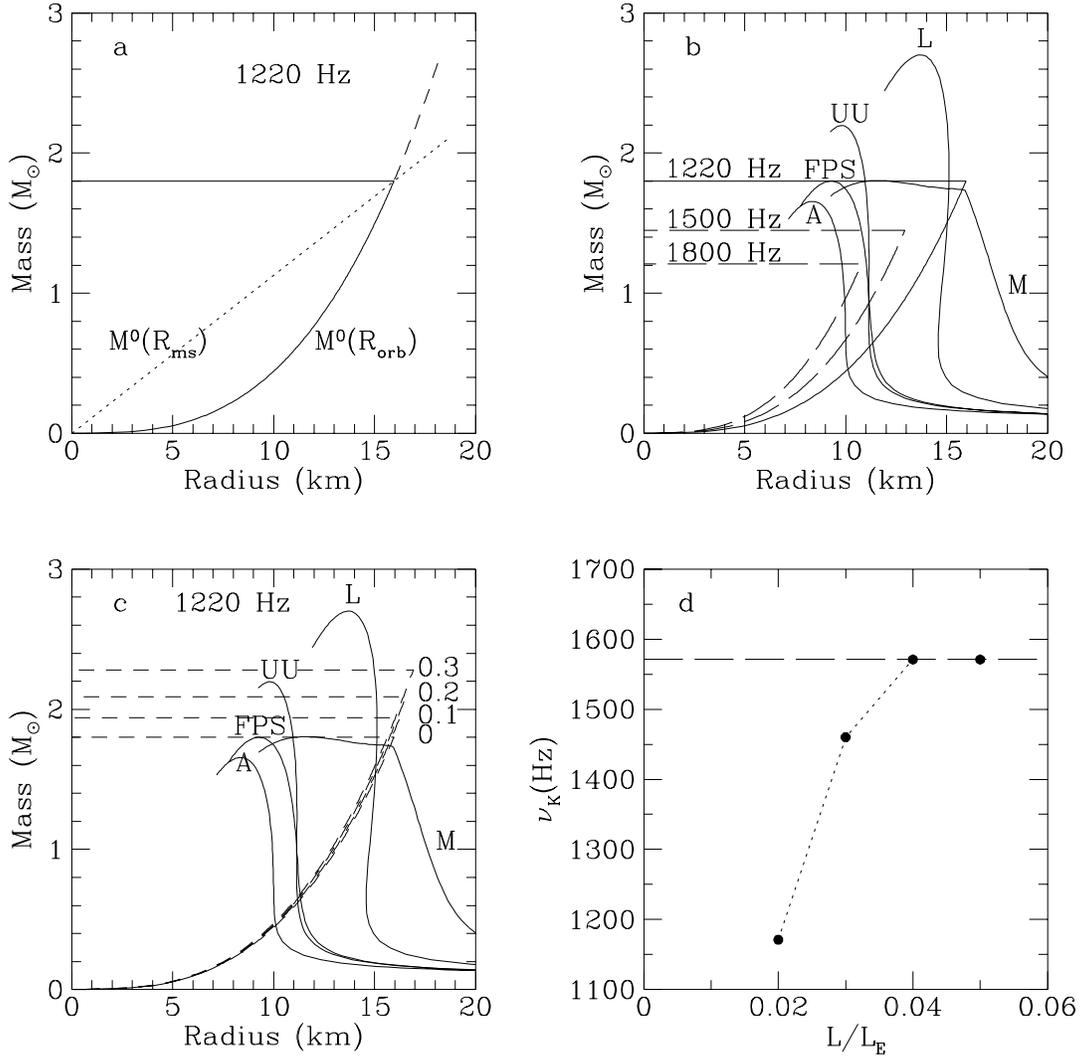,height=6.1truein,width=6.1truein}}
 \caption{
 (a)~Radius-mass plane, showing how bounds on
the mass and radius of a nonrotating neutron
star with $\nu_{\rm QPO2}^\ast = 1220$~Hz can be
constructed. The steps in the construction are
similar for rotating stars.
  (b)~Comparison of the $M$-$R$ relations for
nonrotating neutron stars given by five
representative equations of state (solid
curves) with the regions of the radius-mass
plane allowed for nonrotating stars with three
different circumstellar orbital frequencies.
The region bounded by the heavy solid line is
the region that would be allowed for
4U~1636$-$536 if it were not rotating.
  (c)~Regions allowed for rotating neutron stars
with various values of $j$ and $\nu_{\rm
QPO2}^\ast = 1220$~Hz, when the first-order
effects of the stellar spin are included. The
gas generating the QPO is assumed to be in a
prograde orbit.
  (d)~Characteristic variation of the sonic-point
orbital frequency with accretion luminosity from
fully general relativistic calculations of the
gas dynamics and radiation transport. The
sonic-point orbital frequency increases steeply
with increasing accretion luminosity until it
reaches the frequency of the ISCO, at which
point it stops changing.
 See [2] for further details.
 }
 \end{figure}

The spin of the star affects both the structure
of the star and the spacetime, altering the
stellar mass-radius relation, the frequency of an
orbit of given radius, and the value of $R_{\rm
ms}$ for a given stellar mass. The parameter that
characterizes the importance of these effects is
the dimensionless quantity \hbox{$j \equiv
cJ/GM^2$}, where $J$ and $M$ are the angular
momentum and gravitational mass of the star. For
the spin frequencies $\sim 300~\Hz$ inferred in
the kilohertz QPO sources, $j \sim 0.1 \dash
0.3$. For such small values of $j$, a first-order
treatment is adequate.

To first order in $j$, the frequency of the
prograde orbit at $R_{\rm ms}$ around a star of
given mass $M$ and dimensionless angular
momentum $j$ is (see [5])
 $
 \nu_{\rm K, ms} \approx
   2210\,(1+0.75j)(M_\odot/M)\,\Hz
 $.
 The corresponding upper bounds on the mass and
radius are given implicitly by
 \begin{equation}
   M_{\rm max} \approx
   [1+0.75j(\nu_{\rm spin})]M^\nonrot_{\rm max}
  \quad{\rm and}\quad
   R_{\rm max} \approx
   [1+0.20j(\nu_{\rm spin})]
   R^\nonrot_{\rm max}\;,
 \label{MmaxRmax}
 \end{equation}
 where $j(\nu_{\rm spin})$ is the value of $j$
for the observed stellar spin rate at the maximum
allowed mass for the equation of state being
considered. Figure~1c indicates the first-order
effects of spin rates $\sim 300~\Hz$ on the
allowed region of the $R$-$M$ plane. Our
calculations [2] show that {\it the mass
of the neutron star in \fu{1636$-$536} must be
less than $\sim 2.2\,M_\odot$ and its radius
must be less than $\sim 17$~km}. As just
explained, the precise upper bounds depend on
the equation of state assumed. For further
details, see [2].

If the stellar spin frequency is $\sim 500~\Hz$
or higher, spin affects the structure of the
star as well as the exterior spacetime, which
then differs substantially from the Kerr
spacetime and must be computed numerically for
each assumed equation of state. We have
carried out such computations and
find [6] that if the neutron star is
spinning rapidly, the constraints on the
equation of state are dramatically tightened.

\section*{Innermost Stable Circular Orbit}

Establishing that an observed QPO frequency is
the frequency of the ISCO would be an important
step forward in our understanding of strong-field
gravity and the properties of dense matter,
because it would confirm one of the key
predictions of general relativity in the
strong-field regime and fix the mass of the
neutron star in that source, for each assumed
equation of state. Given the fundamental
significance of the ISCO, it is very important
to establish what would constitute strong,
rather than merely suggestive, evidence that an
ISCO has been detected (for a detailed
discussion of various signatures, see
[2]). Probably the most convincing
signature would be a fairly coherent, kilohertz
QPO with a frequency that reproducibly increases
steeply with increasing accretion rate but then
becomes constant and remains nearly constant as
the accretion rate increases further (see
Fig.~1d). The constant frequency should always
be the same in a given source.

It has been suggested that the similarity of the
highest QPO frequencies seen so far indicates
that ISCOs are being detected [7] and
that the roughly constant frequencies of the
800--900~Hz QPOs seen in \fu{1608$-$52}
[8] and \fu{1636$-$536} [9] during
the first observations of these sources with the
{\em Rossi Explorer} were generated by the beat
of the spin frequency against the frequency of
ISCOs in these sources [10]. This would
imply that the neutron stars in all the
kilohertz QPO sources have masses close to
$2.0\,M_\odot$. However, {\em no strong
signatures of the innermost stable circular
orbit have so far been seen in any of these
sources}. Indeed, more recent observations of
both \fu{1608$-$52} [11] and
\fu{1636$-$536} [12] are inconsistent with
the suggestion that the QPO frequencies seen
initially are related to the frequencies of
ISCOs in these sources. Nevertheless, the
prospects appear excellent for discovering
clear evidence of an ISCO in one or more of the
kilohertz QPO sources in the future.

\end{document}